\documentclass[12pt]{article}

\usepackage[cp1251]{inputenc}
\usepackage[T2A]{fontenc}
\usepackage[russian]{babel}
\usepackage{graphicx}
\usepackage{amsmath}

\def\be{\begin{equation}}
\def\ee{\end{equation}}
\def\pe{\perp}

\def\beq{\begin{eqnarray}}
\def\eeq{\end{eqnarray}}

\def\ve{\varepsilon}
\def\dc{\partial}
\def\bc{\begin{center}}
\def\ec{\end{center}}
\def\bt{\begin{tabular}}
\def\et{\end{tabular}}

\begin{document}

\bc {\bf On the Application of Poincare-Steklov Operators to the Problem of Resonant Scattering in a Cylinder.} \ec

\bc {\bf Delitsyn A.L.} \ec

\bc Higher School of Modern Mathematics MIPT, 1 Klimentovskiy per., Moscow, Russia \\
Russian Biotechnological University (ROSBIOTECH), 11 Volokolamskoe shosse, Moscow, 125080, Russia \ec

\bc Abstract. \ec

The resonant nature of scattering in a waveguide with two barriers is proven in the case of a sufficiently arbitrary deformation of the region between the barriers. The problem is considered as an interior boundary value problem with boundary conditions defined by a Poincare-Steklov operator. A spectral problem is considered whose eigenvalue determines the resonant scattering frequency.

~~

Resonant scattering in a cylinder has been the subject of numerous mathematical works, beginning with \cite{Arseniev1}-\cite{Arseniev4}. In this paper, the problem of resonant wave propagation in a cylinder with two barriers is considered. This problem was considered in the author's papers (\cite{Del1})-(\cite{Del3}) by the partial domain method, and in paper \cite{N1} by the matched asymptotic expansion method. The matched asymptotic expansion method was applied in a number of papers \cite{N1}-\cite{S} to solve various versions of the resonant scattering problem. It should be noted that a very wide range of problems can be investigated by this method. At the same time, the author hesitates to call this method elementary. Direct application of the partial domain method is limited to domains composed of cylinders. At the same time, a distinctive feature of this method is related to the consideration of the scattering problem as an interior boundary value problem with boundary conditions determined by Poincare-Steklov type operators. The goal of this paper is to show that the interior domain separated by barriers, which from a physical point of view represents a resonator, can have a fairly arbitrary, not necessarily cylindrical, geometry. Moreover, the proof, from the author's point of view, is extremely simple.

The scattering problem is considered (Fig. 1).


\begin{figure}[ht]
\centering
\includegraphics[width=105mm,height=75mm]{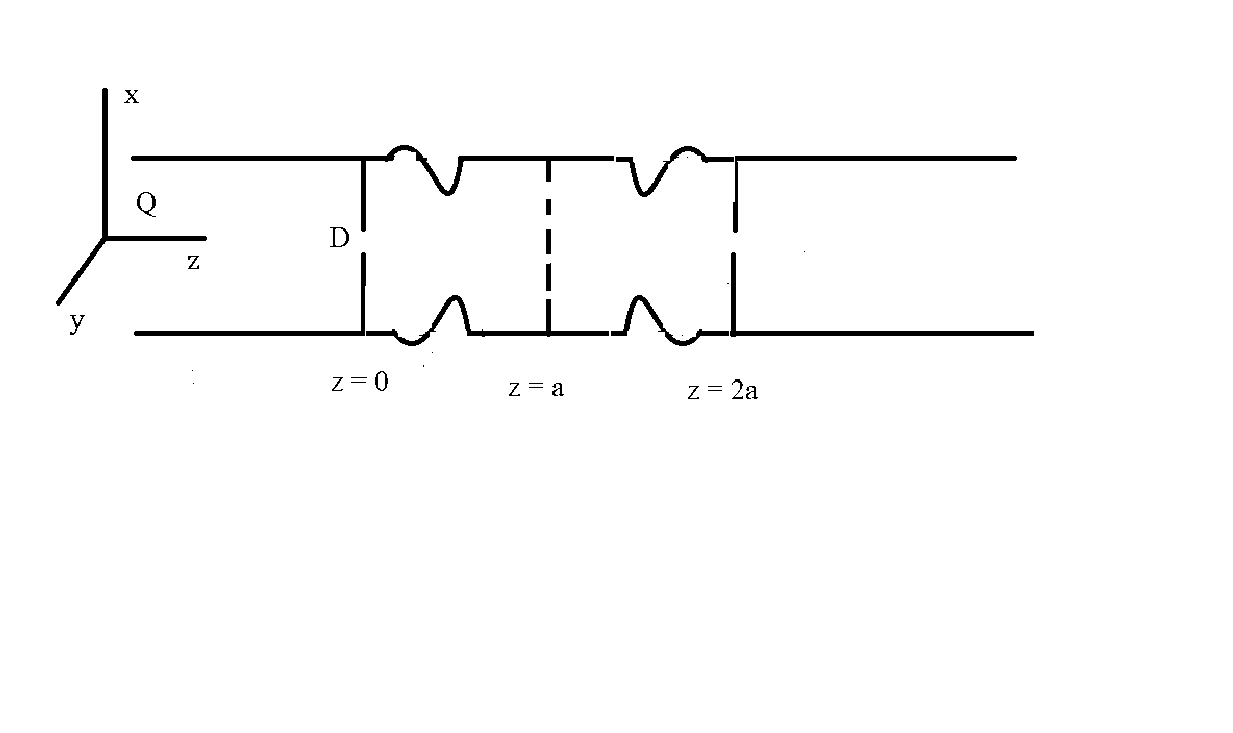}
\caption{Waveguides coupled through a resonator.}
\label{fig2}
\end{figure}

\be \Delta u + k^2 u = 0, \quad \quad (x, y, z) \in Q, \label{1} \ee
\be u|_{\dc Q} = 0 \label{2} \ee
and radiation conditions
\be u = e^{i \gamma_1 z} \psi_1(x, y) + r_1 e^{-i \gamma_1 z} \psi_1(x, y) + \sum \limits_{n=2}^{\infty} r_n e^{\gamma_n z} \psi_n(x, y), z < 0, \label{3} \ee
\be u = t_1 e^{\gamma_1 z} \psi_1(x, y) + \sum \limits_{n=2}^{\infty} t_n e^{-\gamma_n z} \psi_n(x, y), z > 2a, \label{4} \ee
where
$ \psi_n $ are the eigenfunctions, $ \lambda_n $ are the eigenvalues of the Laplace operator with the Dirichlet conditions in the cross section $ \Omega $
$$ - \Delta_{\pe} \psi = \lambda_n \psi_n, \quad \quad (x, y) \in \Omega, \quad \quad \psi_n|_{\dc \Omega} = 0, $$
$$ \gamma_1 = \sqrt{k^2 - \lambda_1}, \gamma_n = \sqrt{\lambda_n - k^2}, n = 2, \dots $$
We consider the case of propagation of a single traveling wave in an infinite cylinder, which corresponds to the condition $ \lambda_1 < k^2 < \lambda_2 $.
The problem is considered in a domain $ Q $ with two planes of symmetry $ z = a $ and $ y = 0 $.
The domain in which the wave propagates is a cylinder with two barriers $ S_i, i = 1, 2 $, located at $ z = 0 $ and $ z = 2a $. The barriers contain identical small apertures $ D_i, i = 1, 2 $, symmetrically located with respect to the plane $ z = a $.
The goal of the work is to prove the existence of a value $ k^2 \in (\lambda_1, \lambda_2) $ such that practically complete transmission, rather than reflection, occurs. However, it is not claimed that the transmission is complete. At the same time, for any given number, there exists an aperture diameter and a value $ k^2 $ for which the reflection coefficient is smaller in absolute value than this number.

This problem was considered in \cite{Del1}-\cite{Del2} for the case where the domain consists of a cylinder with barriers. We will call the region between the barriers $ z = 0, z = 2a $ the resonator. The goal of this paper is to show that, for a certain resonator deformation, a method similar to that used in \cite{Del1}-\cite{Del2} can be applied, and the requirement that the resonator have a cylindrical shape is not essential.

Just as in the work \cite{Del1}-\cite{Del2}, problem (\ref{1})-(\ref{2}) is reduced to the consideration of two problems in the domain $ Q_1 $, distinguished by the condition $ z < a $. For one of which the Dirichlet conditions are used, for the second the Neumann conditions are considered on part of the boundary $ z = a $. In this case, for these problems only the condition on the reflected waves \cite{Del1}-\cite{Del2} remains, and on the boundary $ z = a $ either the condition or $ \frac{\dc u}{\dc z}|_{z=a} = 0 is used. $ The solution to the problem with the Dirichlet condition will be denoted by $ u^D $, and with the Neumann condition by $ u^N $. To prove the resonant nature of scattering for a small aperture diameter, it suffices to prove (\cite{Del2}) that the reflection coefficient of the Neumann problem, denoted by $ r^N $, has a jump from a value close to $ -1 $ to a value equal to $ 1 $ at the resonant value $ k^2 $. The value of the reflection coefficient of the Dirichlet problem, denoted by $ r^D $, remains close to $ -1 $.
Since the solution to the original problem is equal to half the sum of the solutions $ u^D $ and $ u^N $,
$$ u = \frac{1}{2}(u^D + u^N), $$
then the jump in the reflection coefficient $ r^N $ leads to the fact that the reflection coefficient of the original problem $ r $ at resonance will be arbitrarily close to zero, since for the Dirichlet problem the reflection coefficient in the range of $ k^2 $ considered below is close to $ -1 $.
Given that
$$ r^N = (u^N, \psi_1)_{L_2(\Omega)} - 1, $$
the problem reduces to proving that $ (u^N, \psi_1)_{L_2(\Omega)} $ has a jump from a value close to zero to a value equal to $ 2 $.

Thus, we consider the problem for $ u^N $, which we will henceforth denote by $ u $ without the clarifying $ N $. \be \Delta u + k^2 u = 0, \label{21} \ee
\be u|_{\dc Q} = 0, \label{22} \ee
\be \frac{\dc u}{\dc z}|_{z=a} = 0, \label{23} \ee
\be u = e^{i \gamma_1 z} \psi_1(x, y) + r_1 e^{-i \gamma_1 z} \psi_1(x, y) + \sum \limits_{n=2}^{\infty} r_n e^{\gamma_n z} \psi_n(x, y), z < 0, \label{24} \ee

We reformulate problem (\ref{21})-(\ref{24}) as an interior boundary value problem. To do this, we express the coefficients $ r_n $ as
$ r_n = (u, \psi_n)_{L_2(\Omega)} $, differentiate $ u $ with respect to $ z $ for $ z < 0 $, and express the limiting value $ \frac{\dc u}{\dc z} $ as $ z \to 0-0 $ as
\be \frac{\dc u}{\dc z} = -i \gamma_1 (u, \psi_1)_{L_2(\Omega)} \psi_1 + A u + i 2 \gamma_1 \psi_1, \label{25} \ee
where
$$ A u = \sum \limits_{n=2}^{\infty} \gamma_n (u, \psi_n)_{L_2(\Omega)} \psi_n. $$
The operator that associates the function defining the Dirichlet condition $ u $ with the derivative $ \frac{\dc u}{\dc z} $, which defines the Neumann condition, is usually called the Poincare-Steklov operator. When applied to the scattering problem in a cylinder, the boundary conditions for harmonics are also called partial radiation conditions \cite{Sveshnikov}.
We further consider the scattering problem as an interior boundary value problem (\ref{21})-(\ref{23}) with the boundary condition (\ref{25}).

The form of the boundary value problem immediately explains the presence of the resonance effect and the method for determining the resonant frequency.
Let $ k^2 $ be an eigenvalue of the boundary value problem (\ref{21})-(\ref{23}) with the boundary condition
\be \frac{\dc u}{\dc z} = A u. \label{26} \ee
Since problem (\ref{21})-(\ref{23}), (\ref{26}) is homogeneous, then if
\be (u, \psi_1)_{L_2(\Omega)} \ne 0, \label{27} \ee
then, since $ u $ is determined up to a constant, we can assume that $ (u, \psi_1)_{L_2(\Omega)} = 2 $. As a result, $ u $ is a solution to the scattering problem (\ref{21})-(\ref{23}, \ref{25}) and
$$ (u, \psi_1)_{L_2(\Omega)} = 2. $$

Thus, the proof of resonant scattering reduces to the following assertions.

1. It is necessary to prove the existence of an eigenvalue $ k^2 $ of problem (\ref{21})-(\ref{23})-(\ref{26}) close to the eigenvalue of the problem with Dirichlet conditions in the domain $ V $, with the exception of $ z = 0 $, for which the Neumann conditions are imposed.
Furthermore, it is necessary to prove that, under certain restrictions on the domain $ Q $, trapped modes are absent.

2. It is necessary to check that outside the vicinity of the resonant frequency, for a sufficiently small hole diameter, $ (u, \psi_1)_{L_2(D)} $ is small, both for the Dirichlet problem and the Neumann problem.

Let us first prove

{\bf Statement 1}. {\it There exists an eigenvalue $ {k^*}^2 $ of problem (\ref{21})-(\ref{23}), (\ref{26}). Let the hole diameter $ D \to 0 $. Then the eigenvalue $ {k^*}^2 \to k_0^2 $, where $ k_0^2 $ is the eigenvalue of problem (\ref{21})-(\ref{23}) with the Dirichlet condition
\be u|_{D} = 0 \ee
instead of condition (\ref{26}), i.e. resonant frequency of the resonator $ V $.}

Indeed, we introduce a new spectral parameter $ \lambda(k) $ and consider the linear spectral problem:
\be -\Delta u = \lambda(k) u = 0, \label{27} \ee
\be u|_{\dc Q} = 0, \label{28} \ee
\be \frac{\dc u}{\dc z}|_{z=a} = 0, \label{30} \ee
\be \frac{\dc u}{\dc z}|_{z=0} = A u, \label{31} \ee
There exists an eigenvalue $ \lambda(k) $ of the problem (\ref{27}-(\ref{31}). The proof is almost identical to that given in \cite{Del4}. The only difference is that in the operator
$$ A = \sum \limits_{n=2}^{\infty} \gamma_n (u, \psi_n)_{L_2(\Omega)} \psi_n $$
summation is performed from $ 2 $ to $ \infty $, and not from $ 1. $
Just as in that paper, $ \lambda(k) $ is continuous in $ k $ and monotonically decreases on the interval $ [\lambda_1, \lambda_2] $, and on this interval there is an intersection point of $ \lambda(k) $ and $ k^2 $, which implies that the original eigenvalue problem with a nonlinear occurrence of the spectral parameter (\ref{21})-(\ref{23}), (\ref{26}) has an eigenvalue $ {k^*}^2 $.
The eigenvalue $ \lambda(k) $ can be determined variationally
\be \lambda(k) = \inf \limits_{u \in H^1(V), u \ne 0} (u, u)^{-1} ((\nabla u, \nabla u)_{L_2(V)} + (A(k) u, u)). \ee
Let $ {k^N}^2 $ and $ {k^D}^2 $ be the principal eigenvalues of problems (\ref{27})-(\ref{30}) with the Dirichlet or Neumann condition on the hole $ D $ instead of the condition (\ref{31}).

Given the variational definition of the eigenvalues $ {k^N}^2 $ and $ {k^D}^2 $, it holds that
\be {k^N}^2 < {k^*}^2 < {k^D}^2. \ee
Since $ {k^N}^2 \to {k^D}^2 $ as the hole diameter $ d \to 0 $, then $ {k^*}^2 \to {k^D}^2 $.

If the domain $ Q $ is symmetric with respect to the plane $ y = 0 $, the eigenfunction of problem (\ref{21})-(\ref{23}), (\ref{26}) is either even or odd. Since the fundamental eigenfunction is even, the condition (\ref {27}) is satisfied.
Thus, for a value of $ k^2 $ coinciding with $ {k^*}^2 $ close to $ {k^D}^2 $, the solution to the scattering problem satisfies the condition
$$ (u, \psi_1)_{L_2(\Omega)} = 2. $$

{\bf Statement 2}. {\it As the aperture diameter $ d \to 0 $, in some neighborhood of $ {k^D}^2 $ for problems with Dirichlet conditions $ (u^{D}, \psi_1) \to 0 $. It follows that at the resonant frequency $ {k^*}^2 $ there is almost complete passage of the wave through the barriers. If we exclude from the specified neighborhood some small neighborhood to which $ {k^*}^2 $ belongs, then in the resulting range of $ k^2 $ for a problem with the Neumann condition, we have
$ (u^{N}, \psi_1) \to 0 $ as $ d \to 0. $ That is, in this range of $ k^2 $, the transmission coefficient varies from a value close to $ 0 $ to a value arbitrarily close to $ 1 $.}

Let the aperture diameter $ d $ be small enough so that $ \lambda^D - \lambda^N < \ve $.
The solution to the scattering problem satisfies the equation
\be (\nabla u, \nabla u)_{L_2(V)} - k^2 (u, u)_{L_2(V)} + (A u, u))_{L_2(D)} - i \gamma_1 |(u, \psi_1)_{L_2(D)}|^2 = - 2 i \gamma_1 (\psi_1, u)_{L_2(D)}. \ee
It follows that
\be |(u, \psi_1)_{L_2(D)}|^2 = 2 Im (\psi_1, u)_{L_2(D)} \leq 2 |u, \psi_1)_{L_2(D)}|. \ee
As a result,
\be |(u, \psi_1)_{L_2(D)}| < 2.\ee
Besides,
\be (\nabla u, \nabla u)_{L_2(V)} - k^2 (u, u)_{L_2(V)} + (A u, u))_{L_2(D)} = - Im 2 \gamma_1 (\psi_1, u)_{L_2(D)} \leq 2 \gamma_1 |(\psi_1, u)_{L_2(D)}| \leq 4 \gamma_1. \ee
Hence, since
$$ (\nabla u, \nabla u)_{L_2(V)} - k^2 (u, u)_{L_2(V)} + (A u, u))_{L_2(D)} \geq ({k^N}^2 - k^2) ||u||^2_{L_2(V)} \geq ({k^D}^2 - k^2 - \ve) ||u||^2_{L_2(V)}, $$
That
\be ||u||^2_{L_2(V)} \leq \frac{4 \gamma_1}{{k^D}^2 - k^2 - \ve}, \ee
where
\be ||\nabla u||^2_{L_2(V)} \leq 2 + \frac{4 \gamma_1}{{k^D}^2 - k^2 - \ve}. \ee
Since
$$ ||u||_{L_2(D)} \leq C ||u||_{H^1(V)}, $$
then
\be |(u, \psi_1)|_{L_2(D)} \leq ||u||_{L_2(D)}||\psi_1||_{L_2(D)}. \ee
Since $ ||u||_{L_2(D)} $ is bounded and
$$ ||\psi_1||_{L_2(D)} \to 0 $$
as $ d \to 0 $, $ (u, \psi_1)_{L_2(D)} $ is arbitrarily small as $ d \to 0 $.
Similarly, the limit relation for the problem with the Neumann condition holds for $ z = a $.

As a result, we find that this method of proving resonant scattering is not associated with a special type of domain boundary and is applicable to problems with a fairly general form of the resonator.

\end{document}